\documentclass[prd,twocolumn, tightten, aps, nofootinbib, showpacs, preprintnumbers]{revtex4}
\begin{document}
\preprint{USM-TH-216}
\title{Confinement from gluodynamics in curved space-time}
\author{Patricio Gaete} \email{patricio.gaete@usm.cl}
\affiliation{Departamento de F\'{\i}sica, Universidad T\'{e}cnica
Federico Santa Mar\'{\i}a, Valpara\'{\i}so, Chile}
\author{Euro Spallucci}
\email {euro@ts.infn.it} \affiliation{Dipartimento di Fisica
Teorica, Universit\`a di Trieste and INFN, Sezione di Trieste,
Italy}
\date{\today}

\begin{abstract}
We determine the static potential for a heavy quark-antiquark pair
from gluodynamics in curved space-time. Our calculation is done
within the framework of the gauge-invariant, path-dependent,
variables formalism. The potential energy is the sum of a Yukawa and
a linear potential, leading to the confinement of static charges.
\end{abstract}
\maketitle

\section{Introduction}

One of the fundamental issues facing QCD at low energy is a
quantitative description of confinement. We observe at this point
that the distinction between the apparently related phenomena of
screening and confinement is of considerable importance in our
present understanding of gauge theories. In fact, gauge theories
that yield a linear potential are important to particle physics,
since those theories may be used to describe the confinement of
quarks and gluons and be considered as effective theories of QCD. As
is well known, the confinement problem has been fairly well
discussed under a number different aspects, like lattice gauge
theory techniques \cite{Wilson} and non-perturbative solutions of
Schwinger-Dyson's equations \cite{Zachariesen,Depple}. Recently, an
appealing proposal to this problem was made by 't Hooft \cite{'t
Hooft} which includes a linear term in the dielectric field that
appears in the energy density.

In this connection it becomes of interest, in particular, to recall
that QCD at the classical level possesses scale invariance which is
broken by quantum effects. We further observe that this phenomenon
can be mathematically described by formulating classical
gluodynamics in a curved space-time with non vanishing cosmological
constant \cite{Levin1,Levin2}. Correspondingly, an effective low
energy Lagrangian for gluodynamics which describes semi-classical
vacuum fluctuations of gluon field at large distances is obtained
\cite{Levin1,Levin2}.

In the light of the above observations, we also mention that the
Cornell potential \cite{Eichten} which simulates the features of
$QCD$ is given by
\begin{equation}
V =  - \frac{{\kappa}}{r} + \frac{r}{{a^2 }}, \label{Cornell}
\end{equation}
where $a$ is a constant with the dimensions of length. In accordance
with the 't Hooft  proposal, confinement is associated to the
appearance of a linear term in the dielectric field ${\bf D}$ (that
dominates for low $|{\bf D}|$) in the energy density \cite{'t
Hooft}:
\begin{equation}
U({\bf D}) = \rho _{str} |{\bf D}|, \label{'t Hooft}
\end{equation}
the proportionality constant being the coefficient of the linear
potential, that is, $\rho _{str}  \approx \frac{1}{{a^2 }}$. Hence
we see that the confinement phenomena breaks the scale invariance as
the Cornell potential explicitly shows by introducing the scale $a$.

Very recently \cite{GaeteG,GaeteS}, we have approached the
connection between scale symmetry breaking and confinement in a
phenomenological way using the gauge-invariant but path-dependent
variables formalism, which is alternative to the Wilson loop
approach. More specifically, we have shown the appearance of the
Cornell potential (\ref{Cornell}) as well as the 't Hooft relation
(\ref{'t Hooft}) after spontaneous breaking of scale invariance in
both Abelian and non-Abelian cases. Certainly, this study gives us
an opportunity to compare our results with that of gluodynamics in
curved space-time \cite{Levin1,Levin2}. To this end, we will work
out the static potential for the theory under consideration along
the lines of Ref. \cite{GaeteG,GaeteS}. As a result, it is found
that the potential energy is the sum of a Yukawa and a linear
potential, leading to the confinement of static charges. This static
potential clearly shows the key role played by the quantum
fluctuations. In general, this picture agrees qualitatively with
that encountered in our previous phenomenological model
\cite{GaeteS}. It is important to realize that the gluodynamics in
curved space-time studied here, compared with our previous model
which includes a $ \sqrt{ F_{\mu\nu}^{a}F^{a\mu\nu}}$ term coupled
to the Yang-Mills Lagrangian density, involves the contribution of
quantum fluctuations. Hence we see that our phenomenological model
incorporates automatically the contribution of these quantum
fluctuations to the vacuum of the model. In this way we establish a
new correspondence between these two non-Abelian effective theories.
The above connections are of interest from the point of view of
providing unifications among diverse models as well as exploiting
the equivalence in explicit calculations.

\section{Interaction energy}

We now examine the interaction energy between static point-like
sources for gluodynamics in curved space-time. To do this, we will
compute the expectation value of the energy operator $H$ in the
physical state $|\Phi\rangle$ describing the sources, which we will
denote by $ {\langle H\rangle}_\Phi$. However, before going to the
derivation of the interaction energy, we will describe very briefly
the model under consideration. The initial point of our analysis is
the dilaton effective lagrangian coupled to gluodynamics
\cite{Levin1}:
\begin{eqnarray}
{\cal L} &=& \frac{{\left| {\varepsilon _V } \right|}}{{m^2
}}\frac{1}{2}e^{{\raise0.7ex\hbox{$\chi $} \!\mathord{\left/
 {\vphantom {\chi  2}}\right.\kern-\nulldelimiterspace}
\!\lower0.7ex\hbox{$2$}}} \left( {\partial _\mu  \chi } \right)^2  +
\left| {\varepsilon _V } \right|e^\chi  \left( {1 - \chi } \right) \nonumber \\
&-& e^\chi  \left( {1 - \chi } \right)\frac{1}{4}F_{\mu \nu }^a F^{a\mu
\nu }, \label{gluo5}
\end{eqnarray}
where the real scalar field $\chi$ of mass $m$ represents the
dilaton, and $- \left| {\varepsilon _V } \right|$ is the vacuum
energy density.

By expanding near the $\chi=0$, expression (\ref{gluo5}) then
becomes
\begin{eqnarray}
{\cal L} &=&  - \frac{1}{4}F_{\mu \nu }^a F^{a\mu \nu }  +
\frac{{\left| {\varepsilon _V } \right|}}{{2 m^2 }}\left( {\partial
_\mu \chi } \right)^2  \nonumber \\
&-& \chi \left( {\left| {\varepsilon _V } \right| -
\frac{1}{4}\left( {F_{\mu \nu }^a } \right)^2 } \right).
\label{gluo10}
\end{eqnarray}
Following our earlier procedure \cite{GaeteG,GaeteS}, integrating
out the $\chi$ field induces an effective theory for the $A^a_{\mu}$
field. Once this is done, we arrive at the following effective
Lagrangian density:
\begin{eqnarray}
{\cal L}_{eff}  &=&  - \frac{1}{4}F_{\mu \nu }^a \left( {1 +
\frac{{m^2 }}{\Delta }} \right)F^{a\mu \nu } \nonumber \\
&+& \frac{{m^2 }}{{32\left| {\varepsilon _V } \right|}}\left(
{F_{\mu \nu }^a } \right)^2 \frac{1}{\Delta }\left( {F_{\mu \nu }^a
} \right)^2. \label{gluon15}
\end{eqnarray}
Next, in order to linearize this theory , we introduce the auxiliary
field $\phi$. It follows that the expression (\ref{gluon15}) can be
rewritten as
\begin{eqnarray}
{\cal L}_{eff}  &=&  - \frac{1}{4}F_{\mu \nu }^a \left( {1 +
\frac{{m^2 }}{\Delta }} \right)F^{a\mu \nu }  + \frac{1}{2}\left(
{\partial _\mu  \phi } \right)^2  \nonumber \\
&-& \frac{1}{4}\frac{m}{{\sqrt
{\left| {\varepsilon _V } \right|} }}\phi \left( {F_{\mu \nu }^a }
\right)^2. \label{gluon20}
\end{eqnarray}
Once again, by expanding about $\phi=\phi_0$, we then obtain
\begin{equation}
{\cal L}_{eff}  =  - \frac{1}{4}F_{\mu \nu }^a \frac{1}{\varepsilon
}\left( {1 + \frac{{\varepsilon m^2 }}{\Delta }} \right)F^{a\mu \nu
}, \label{gluon25}
\end{equation}
where $\frac{1}{\varepsilon } \equiv 1 + \frac{m}{{\sqrt {\left|
{\varepsilon _V } \right|} }}\phi _0$.

To obtain the corresponding Hamiltonian, we must carry out the
quantization of this theory. The Hamiltonian analysis starts with
the computation of the canonical momenta $\Pi ^{a\mu } = -
\frac{1}{\varepsilon }\left( {1 + \frac{{\varepsilon m^2 }}{\Delta
}} \right)F^{a0\mu } $, and one immediately identifies the primary
constraint $\Pi^{a0}=0$ and $\Pi ^{ai }  =  - \frac{1}{\varepsilon
}\left( {1 + \frac{{\varepsilon m^2 }}{\Delta }} \right)F^{a0i }$.
Standard techniques for constrained systems then lead to the
following canonical Hamiltonian:
\begin{eqnarray}
H_C &=& \int {d^3 x} \left\{ {\frac{\varepsilon }{2}\Pi ^{ai} \left(
{1 + \frac{{\varepsilon m^2 }}{\Delta }} \right)^{ - 1} \Pi ^{ai} }
\right\} \nonumber \\
&+& \int {d^3 x} \left\{ {\frac{1}{{4\varepsilon }}F_{ij}^a \left(
{1 + \frac{{\varepsilon m^2 }}{\Delta }} \right)F^{aij} } \right\}
\nonumber \\
&+& \int {d^3 x} \left\{ {\Pi ^{ai} \left( {\partial _i A_0^a +
gf^{abc} A_0^c A_i^b } \right)} \right\}. \label{gluon30}
\end{eqnarray}
The persistence of the primary constraints $\Pi^{a0}\approx0$ leads
to the following secondary constraints $\Gamma ^{a \left( 1 \right)}
\left( x \right) \equiv
\partial _i \Pi ^{ai}  + gf^{abc} A^{bi} \Pi _i^c \approx 0$. It is
easy to check that there are no further constraints, and that the
above constraints are first class. Therefore, the extended
Hamiltonian that generates translations in time then reads $H = H_C
+ \int d x \left( {c_0^{a} (x)\Pi_0^{a} (x) + c_1^{a} (x)\Gamma ^{a
\left( 1 \right)} \left( x \right)} \right)$, where $c_0^{a}(x)$ and
$c_1^{a}(x)$ are the Lagrange multipliers. Moreover, it follows from
this Hamiltonian that $ \dot{A}_0^{a} \left( x \right) = \left[
{A_0^{a} \left( x \right),H} \right] = c_0^{a} \left( x \right)$,
which are arbitrary functions. Since $\Pi^{0a} = 0 $, neither $
A^{0a}$ nor $\Pi^{0a}$ are of interest in describing the system and
may be discarded from the theory. The Hamiltonian then takes the
form
\begin{eqnarray}
H &=& \int {d^3 x} \left\{ {\frac{\varepsilon }{2}{\bf \Pi} ^{a}
\left( {1 + \frac{{\varepsilon m^2 }}{\Delta }} \right)^{ - 1} {\bf
\Pi} ^{a} }
\right\} \nonumber \\
&+& \int {d^3 x} \left\{ {\frac{1}{{2\varepsilon }}{\bf B}^a \left(
{1 + \frac{{\varepsilon m^2 }}{\Delta }} \right){\bf B}^{a} }
\right\}
\nonumber \\
&+& \int {d^3 x} \left\{{ c^a  \left( x \right) \left( {\partial _i
\Pi^{ai} + gf^{abc} A^{bi}\Pi _i^c } \right)} \right\}.
\label{gluon35}
\end{eqnarray}
where $c^a  \left( x \right) = c_1^{a} \left( x \right) - A_0^{a}
\left( x \right)$. According to the usual procedure we introduce a
supplementary condition on the vector potential such that the full
set of constraints becomes second class. A particularly useful and
interesting choice is given by \cite{Gaete1}:
\begin{equation}
\Gamma^{a\left( 2 \right)} \left( x \right) = \int\limits_0^1
{d\lambda } \left( {x - \xi } \right)^i A_i^{\left( a \right)}
\left( {\xi  + \lambda \left( {x - \xi } \right)} \right) \approx 0,
\label{gluon40}
\end{equation}
where  $\lambda$ $(0\leq \lambda\leq1)$ is the parameter describing
the spacelike straight path $ x^i  = \xi ^i  + \lambda \left( {x -
\xi } \right)^i $, on a fixed time slice. Here $ \xi $ is a fixed
point (reference point), and there is no essential loss of
generality if we restrict our considerations to $ \xi ^i=0 $. As a
consequence, the only nontrivial Dirac bracket is
\begin{eqnarray}
\left\{ {A_i^a \left( x \right),\Pi ^{bj} \left( y \right)}
\right\}^ *   = \delta ^{ab} \delta _i^j \delta ^{(3)} \left( {x -
y} \right) \nonumber \\
- \int\limits_0^1 {d\lambda } \left( {\delta ^{ab} \frac{\partial
}{{\partial x^i }} - gf^{abc} A_i^c \left( x \right)} \right)x^j
\delta ^{(3)} \left( {\lambda x - y} \right). \label{gluon45}
\end{eqnarray}
In passing we note the presence of the last term on the right-hand
side which depends on $g$.

Now we move on to compute the interaction energy between pointlike
sources in the theory under consideration, where a fermion is
localized at the origin $ {\bf 0}$ and an antifermion at $ {\bf y}$.
As mentioned before, in order to accomplish this purpose we will
calculate the expectation value of the energy operator $ H$ in the
physical state $|\Phi\rangle $. From our above discussion we see
that $\left\langle H \right\rangle _\Phi$ reads
\begin{equation}
\left\langle H \right\rangle _\Phi   = \frac{\varepsilon }{2}
 tr\left\langle \Phi  \right|\int {d^3 x}  \Pi ^{ai} \left( {1 +
\frac{{\varepsilon m^2 }}{\Delta }} \right)^{ - 1} \Pi ^{ai} \left|
\Phi  \right\rangle. \label{gluon50}
\end{equation}

At this stage we recall that the physical state can be written as \cite
{Gaete1},
\begin{equation}
\left| \Phi  \right\rangle  = \overline \psi  \left( {\bf y}
\right)U\left( {{\bf y},{\bf 0}} \right) \psi \left( {\bf 0}
\right)\left| 0 \right\rangle, \label{gluon55}
\end{equation}
where
\begin{equation}
U\left( {{\bf y},{\bf 0}} \right) \equiv P\exp \left( {ig\int_{\bf
0}^{\bf y} {dz^i A_i^a \left( z \right)T^a } } \right).
\label{gluon60}
\end{equation}
As before, the line integral is along a spacelike path on a fixed
time slice, $P$ is the path-ordering prescription and $\left|
0\right\rangle$ is the physical vacuum state. As in \cite{Gaete1},
we again restrict our attention to the weak coupling limit.

From the above Hamiltonian analysis the static potential is divided
into two parts: an Abelian part $V^{\left( 1 \right)}$ (proportional
to $C_{F}$) and a non-Abelian part $V^{\left( 2 \right)}$
(proportional to the combination $C_{F}C_{A}$). Thus $\left\langle H
\right\rangle _\Phi$ takes the form
\begin{equation}
\left\langle H \right\rangle _\Phi   = \left\langle H \right\rangle
_0  + V^{\left( 1 \right)} + V^{\left( 2 \right)}, \label{gluon65}
\end{equation}
where $\left\langle H \right\rangle _0  = \left\langle 0
\right|H\left| 0 \right\rangle$. The $V^{\left( 1 \right)}$ and
$V^{\left( 2 \right)}$ terms are given by
\begin{eqnarray}
V^{\left( 1 \right)} &=&  - \frac{{g^2 }}{2}\varepsilon tr\left(
{T^a T^a } \right)\int {d^3 x} \int_{\bf 0}^{\bf y} {dz_i^ {\prime}
} \delta ^{\left(
3 \right)} \left( {{\bf x} - {\bf z}^ {\prime}  } \right) \nonumber \\
&\times& \frac{1}{{\nabla _x^2-\varepsilon m^2 }}\nabla
_x^2\int_{\bf 0}^{\bf y} {dz^i } \delta^{\left( 3 \right)} \left(
{{\bf x} - {\bf z}} \right), \label{gluon70}
\end{eqnarray}
and
\begin{eqnarray}
V^{\left( 2 \right)}  &=& \frac{\varepsilon }{2}{\cal N}\left\langle
0 \right|\int_{\bf 0}^{\bf y} {dz^{ {\prime } k} A_k^c } \left( {z^
{\prime } } \right)z^{ {\prime } i} \int_0^1 {d\lambda } \delta
^{\left( 3 \right)} \left( {\lambda {\bf z}^ {\prime }
- {\bf x}} \right)  \nonumber \\
&\times&\frac{1}{{\nabla _x^2  - \varepsilon m^2 }}\nabla _x^2
\int_{\bf 0}^{\bf y} {dz^l } A_l^e \left( z \right) \nonumber \\
&\times&z^i \int_0^1 {d\beta } \delta ^{\left( 3 \right)} \left(
{\beta {\bf z} - {\bf x}} \right)\left| 0 \right\rangle,
\label{gluon75}
\end{eqnarray}
where ${\cal N} = g^4 tr\left( {f^{abc} T^b f^{ade} T^d } \right)$
and the integrals over $z^{i}$ and $z^{\prime }_{i}$ are zero except
on the contour of integration.

Here,  the $V^{\left( 1 \right)}$ term gives a Yukawa-type potential
plus self-energy terms. In effect, expression (\ref{gluon70}) can
also be written as
\begin{equation}
V^{\left( 1 \right)}  = \frac{{g^2 }}{2}\varepsilon tr\left( {T^a
T^a } \right)\int_{\bf 0}^{\bf y} {dz_i^ {\prime }  } \partial
_i^{z^ {\prime } } \int_{\bf 0}^{\bf y} {dz^i } \partial _z^i
G\left( {{\bf z},{\bf z}^{\prime } } \right), \label{gluon80}
\end{equation}
where $G$ is the Green function
\begin{equation}
G\left( {{\bf z},{\bf z}^ {\prime}  } \right) = \frac{1}{{4\pi
}}\frac{{e^{ - m\sqrt \varepsilon  |{\bf z} - {\bf z}^ {\prime}  |}
}}{{|{\bf z} - {\bf z}^ {\prime} |}}. \label{gluon85}
\end{equation}
Employing Eq.(\ref{gluon85}) and remembering that the integrals over
$z^{i}$ and $z^{\prime }_{i}$ are zero on the contour of
integration, expression (\ref{gluon80}) reduces to the familiar
Yukawa potential  after subtracting the self-energy terms. In other
words,
\begin{equation}
V^{\left( 1 \right)}  =  - \frac{{g^2 }}{{4\pi }}\varepsilon
C_{F}\frac{{e^{ - m\sqrt \varepsilon  L} }}{L}, \label{gluon90}
\end{equation}
where $\left| {\bf y} \right| \equiv L$ and $tr(T^{a}T^{a})=C_{F}$.
We also recall that $\varepsilon \equiv \left( {1 + \frac{m}{{\sqrt
{\left| {\varepsilon _V } \right|} }}\phi _0 } \right)^{ - 1}$.

We now turn our attention to the calculation of the $V^{\left( 2
\right)}$ term, which is given by
\begin{eqnarray}
V^{\left( 2 \right)}  &=& \frac{\varepsilon }{2}{\cal N}^ {\prime }
\int_{\bf 0}^{\bf y} {dz^l } \int_{\bf 0}^{\bf y} {dz^{ {\prime } k}
} D_{lk} \left(
{{\bf z},{\bf z}^ {\prime } }\right) \nonumber \\
&\times&\int_{\bf 0}^{{\bf z}^ {\prime }} {du^i } \int_{\bf 0}^{\bf
z} {dv^i } \left( { - \nabla ^2 } \right)_{\bf u} G\left( {{\bf
u},{\bf v}} \right), \label{gluon95}
\end{eqnarray}
where ${\cal N}^ {\prime }   = g^4 tr\left( {f^{abc} T^b f^{adc} T^d
} \right)$, and $G$ is the Green function. Here $D_{lk} \left( {{\bf
z},{\bf z}^ {\prime } }\right)$ stands for the propagator, which is
diagonal in color space and taken in an arbitrary gauge.

It is appropriate to observe here that the above term is similar to
the one found in non-Abelian axionic electrodynamics
\cite{GaeteSpall}. Nevertheless, in order to put our discussion into
context it is useful to summarize the relevant aspects of the
analysis described previously \cite{GaeteSpall}. In effect, as was
observed in Ref. \cite{GaeteSpall}, by writing the Green function in
momentum space
\begin{equation}
G({\bf u},{\bf v}) = \int {\frac{{d^3 k}}{{\left( {2\pi } \right)^3
}}} \frac{{e^{i{\bf k} \cdot \left( {{\bf u} - {\bf v}} \right)}
}}{{{\bf k}^2 + \varepsilon m^2 }}, \label{gluon100}
\end{equation}
expression (\ref{gluon95}) reduces to
\begin{eqnarray}
V^{\left( 2 \right)}  &=& \frac{\varepsilon }{2}{\cal N}^ {\prime
}\int_{\bf 0}^{\bf y} {dz^l
} \int_{\bf 0}^{\bf y} {dz^{ {\prime } k} } D_{lk} \left( {{\bf z},
{\bf z}^ {\prime }  } \right) \nonumber \\
&\times& \int_{\bf 0}^{{\bf z}^ {\prime } } {du^i } \int_{\bf
0}^{\bf z} {dv^i } \int {\frac{{d^3 k}}{{\left( {2\pi } \right)^3
}}} {\bf k}^2 \frac{{e^{i{\bf k} \cdot \left( {{\bf u} - {\bf v}}
\right)} }}{{{\bf k}^2 + \varepsilon m^2 }} .\label{gluon105}
\end{eqnarray}
Following our earlier procedure  \cite{GaeteSpall}, equation
(\ref{gluon105}) can also be written as
\begin{equation}
V^{\left( 2 \right)}  = \frac{g^{4}}{{8\pi }} \varepsilon C_{F}C_{A}
{\sigma}\int_{\bf 0}^{\bf y} {dz^l } \int_{\bf 0}^{\bf y} {dz^{
{\prime } k} } \left| {\bf z} \right|D_{lk} \left( {{\bf z},{\bf z}^
{\prime }  } \right), \label{gluon110}
\end{equation}
where ${\sigma} \equiv \left[ {\Lambda ^2  - \varepsilon m^2 \ln
 \left( {1 + \frac{{\Lambda ^2 }}{{\varepsilon m^2 }}} \right)} \right] $,
and  $\Lambda$ is a cutoff. As has been shown previously
\cite{GaeteSpall}, we choose $D_{lk} \left( {\bf z},{\bf z}^ {\prime
}   \right)$ in the Feynman gauge. As a consequence, expression
(\ref{gluon110}) then becomes
\begin{equation}
V^{\left( 2 \right)}  = \frac{g^{4} }{{8\pi }}\varepsilon C_{F}C_{A}
{\sigma} L. \label{gluon115}
\end{equation}

From equations (\ref{gluon90}) and (\ref{gluon115}), the
corresponding static potential for two opposite charges located at
${\bf 0}$ and ${\bf y}$ may be written as
\begin{equation}
V= - \frac{{g^2 }}{{4\pi }}\varepsilon
C_{F}\frac{{e^{ - m\sqrt \varepsilon  L} }}{L} +\frac{g^{4} }{{8\pi }}
\varepsilon C_{F}C_{A} {\sigma} L, \label{gluon120}
\end{equation}
where $\left| {\bf y} \right| \equiv L$, and $\varepsilon \equiv
 \left( {1 + \frac{m}{{\sqrt {\left| {\varepsilon _V } \right|} }}
 \phi _0 } \right)^{ - 1}$.
This potential displays the conventional screening part, encoded in
the Yukawa potential, and the linear confining potential. It is
worthwhile noticing that the result (\ref{gluon120}) leads to a
Coulomb-type potential in the limit of large $m$. This then implies
that expression (\ref{gluon120}) has the Cornell form.

\section{Final remarks}

We have studied the equivalence between two non-Abelian effective
theories. To this end we have computed the static potential for a
QCD effective theory which represents propagation and interaction of
gluons and dilaton. As a consequence of this the potential energy is
the sum of a Yukawa and a linear potential, leading to the
confinement of static charges. In a general perspective, this
picture agrees qualitatively with that encountered in our previous
phenomenological model \cite{GaeteS}. We also mention that this
result is in agreement with the studies of Ref. \cite{Dick}.

In this way we have provided a new connection between effective
models. The above analysis reveals the key role played by the
quantum fluctuations in order to obtain confinement.

\section{ACKNOWLEDGMENTS}

 Work supported in part by FONDECYT (Chile) grant 1050546.

\end{document}